\documentclass{article}

\usepackage{graphicx}
\usepackage{bm}
\usepackage[intlimits]{amsmath}

\renewcommand{\d}{\partial}

\newcommand{\q}{{\bf q}}

\newcommand{\p}{{\bf p}}

\newcommand{\rh}{\varrho}
\newcommand{\ep}{\varepsilon}

\newcommand{\exv}[1]{\left\langle{\,#1\,}\right\rangle}

\newcommand{\Tr}{\mathop{\textrm{Tr}}}

\newcommand{\pint}[2]{{\int\!\frac{d^{#1}#2}{(2\pi)^#1}\,}}

\newcommand{\ket}[1]{\left|{#1}\right\rangle}
\newcommand{\nn}{\nonumber\\}

\title{Viscosity to entropy density ratio: violation of the lower bound at
  small temperatures}

\author{A. Jakovac\footnote{on leave from Institute of Physics, BME
 Technical University, H-1521 Budapest, Hungary, jakovac@esr.phy.bme.hu}
\\{\it\small Department of Theoretical Physics, University of Wuppertal,}
 \\{\it\small Gau$\beta$strasse 20, Wuppertal 42119, Germany}
}

\date{March 4, 2009}

\begin{document}

\maketitle

\begin{abstract}
  We show that in theories where the lowest energy excitations are not
  quasiparticles but they form a continuum, the shear viscosity to entropy
  density ratio goes to zero as the temperature goes to zero. In these
  theories therefore there is no lower bound for the shear viscosity to
  entropy density ratio, in contrast to the predictions coming from the
  AdS/CFT correspondence.
\end{abstract}


\section{Introduction}

Experimental evidences from RHIC suggest that the plasma formed after heavy
ion collisions is very close to a perfect fluid. The measured large value of
$v_2$ \cite{Adler:2003kt, Adams:2003am} indicates a streaming with very small
viscosity \cite{Shuryak:2003xe,Teaney:2003kp}: fits to the $v_2$ values
suggest an upper bound $\eta/s<0.16$ \cite{Romatschke}.

There is a tremendous effort in the recent literature to give an account for
such small values of viscosity from the theory side. Perturbation theory a
priori performs an expansion around the ideal gas limit, where $\eta=\infty$,
and so the weak coupling expansion usually yields large $\eta/s$ ratios
proportional to $1/g^4\ln 1/g$
\cite{Arnold:2000dr,Arnold:2003zc,Arnold:2006fz,Gagnon:2007qt}. Still there is
hope to access experimentally acceptable ratios using higher order processes in
Boltzmann equations \cite{Xu:2007ns,Xu:2007jv}. There exist analytic methods
beyond weak coupling perturbation theory, like large $N_f$ expansion
\cite{Aarts:2005vc} or strong coupling expansion in pure Yang-Mills theories
\cite{Jakovac:2008ft}.

One can use Monte Carlo simulations to calculate energy-momentum tensor
correlators, which is related to the viscosity via an integral equation. From
this one can extract the viscosity for pure Yang-Mills theories
\cite{Karsch:1986cq,Meyer:2007ic,Meyer:2008dq}. These results are, however,
not fully reliable yet \cite{Petreczky:2007js}.

A method proved to be very useful in studying the viscosity to entropy ratio
is based on the AdS/CFT correspondence: in ${\cal N}=4$ super Yang-Mills
theory at strong t'Hooft coupling $\lambda$ and at large $N_c$ one finds
$\eta/s=1/4\pi$ \cite{Policastro:2001yc}. In lot of other gravity models we
find the same result \cite{Myers:2008me}. One can also calculate the
$1/\lambda$ corrections \cite{Myers:2008yi}, which turned out to be
positive. It is very natural then to expect, that there is a lower bound to
the viscosity to entropy density ratio in the conformal case. One can argue
that this may be true for other models, too \cite{Kovtun:2004de}: since the
viscosity over entropy ratio parametrically is proportional to $E\tau$, where
$E$ is the quasiparticle energy, $\tau$ is its lifetime, we expect that
Heisenberg uncertainty relation constrains this ratio. The value of the lower
bound is expected to be given by the conformal case, $1/4\pi$.

There are, however, caveats in the argumentation, in two aspects. One is the
gravity side: in higher derivative gravity models the ratio can be smaller
than $1/4\pi$. In the construction of Ref. \cite{Kats:2007mq} the lower bound
in the unitary regime is $16/25\times 1/4\pi$ \cite{Brigante:2008gz}. There
are other string theory constructions, where one can violate the $1/4\pi$
lower bound \cite{Buchel:2008vz}. Recently there are several gravitational
models, where the value of the lower bound is challenged further
\cite{Cai:2009zv,Neupane:2009zz}. This may mean, that in a generic higher
derivative gravity model there is no lower bound at all.

The other caveat comes from the particle physics side: large number of
particles can lead to a large Gibbs mixing entropy, while the viscosity is
constant \cite{Cohen:2007qr,Son:2007xw,Cohen:2007zc,Aarts:2005vc}, and so the
$\eta/s$ ratio approaches zero. In \cite{Cherman:2007fj} it is shown that the
large number of species can be in the same quantum channel as excited states.

In quantum field theories the density of states generally contains a
continuous part, which can be considered as the generalization of the large
number of excited states. In these models under certain conditions we may
expect the violation of the $\eta/s$ bound. This was observed in strongly
coupled pure Yang-Mills theories \cite{Jakovac:2008ft}: when going to zero
temperature the viscosity to entropy density ratio approached zero. From the
details of the calculation one finds that this behavior is the consequence of
that that in the spectrum there are no quasiparticles, only a continuum above
a mass gap.

In this paper we show that this behavior is a common feature of a large class
of relativistic quantum field theories, where the lowest lying excitations of
the energy spectrum form a continuum instead of a discrete set
(quasiparticles). As opposed to systems with quasiparticle excitations (Bose
gases or liquids), these bosonic systems are non-Bose gases (liquids). In
these models the $\eta/s$ ratio approaches zero when the temperature goes to
zero. In fact the argumentation is based on kinematics, and therefore it is
even more generic: we should find this behavior for all transport
coefficients. This result may suggest that these theories have, if they have
at all, a nonlocal dual gravity model.

Before the actual calculation we note that formally all relativistic theories
at zero chemical potential are superfluids, since the spectral function must
be zero below the light cone. It does not mean that the viscosity goes to zero
at zero temperature: in fact also in superfluid $^4$He the shear viscosity is
divergent \cite{Maris}. As we will show the $T\to0$ limit of $\eta/s$ can be
very different for different models: it can be infinite (massive
quasiparticles), finite (conformal case) or zero (non-Bose liquids).

\section{General setup}

To start we consider a self-adjoint operator $A(x)$, and study its correlator
at low temperatures. We use the standard technique \cite{LeBellac:1996} to
write it with the help of matrix elements of $A(0)$. To this end we choose a
basis in the Hilbert space and write at finite temperature
\begin{eqnarray}
  && C_A(x) = \exv{[A(x),A(0)]} =\nn&&
  =\frac1Z \sum_{n,m} \biggl[ \Tr\exv{n\left| e^{-\beta H}
      A(x)\right|m}\exv{m\left|A(0)\right|n}- \nn&& \hspace*{3em} 
  \Tr\exv{m\left| e^{-\beta H}
      A(0)\right|n}\exv{n\left|A(x)\right|m} \biggr]
\end{eqnarray} 
where $Z = \sum_{n\in{\cal N}} \Tr \, e^{-\beta H}$. For the states we assume
that they are represented by the four-momentum of the state $p=(p_0,\p)$, and
other quantum numbers ${\cal Q}$, so we have $\ket{n} = \ket{p,{\cal
    Q}}$. Note, that $p_0$ and $\p$ are not connected in general by a
dispersion relation, not yet the possible values of $p_0$ form a discrete
set. To describe the generic situation we introduce the density of states of a
given quantum channel $\rh_{\cal Q}(p)$, with the definition
\begin{eqnarray}
  \sum_n \exv{n\left|\dots\right|n} =
  \sum_{\cal Q}\!\! \pint 4p\, \rh_{\cal Q}(p)\exv{p,{\cal Q}\left| \dots
    \right|p,{\cal Q}}.
\end{eqnarray}
We will also assume that the normalization of the states are done for unit
volume -- then all calculated quantities are densities. We emphasize here that
the volume of the system is infinite, only the normalization volume of the
states is fixed!

We can write $A(x) =e^{iP^\mu x_\mu} A(0) e^{-iP^\mu x_\mu}$ with the use of
the generator of space-time translation, and so $\exv{q,{\cal Q}\left|
    A(x)\right|p,{\cal P}} = e^{-i(p-q)x} \exv{q,{\cal Q}\left|
    A(0)\right|p,{\cal P}}$. We also perform Fourier transformation, and find
finally
\begin{eqnarray}
  C_A(k) =\frac1Z \sum_{{\cal Q},{\cal P}}&&\pint4q\frac{d^4p}{(2\pi)^4}\,
  \rh_{\cal Q}(q)\rh_{\cal P}(p)\,e^{-\beta q_0} \times\nn&&\times
  \left(1 -e^{-\beta
      k_0}\right)(2\pi)^4\delta(k+q-p)\left|\exv{q,{\cal
        Q}\left|A(0)\right|p,{\cal P}}\right|^2\!\!.
\end{eqnarray}
This form demonstrates also that $C(k)>0\,\forall k_0>0$, and $C(-k)=-C(k)$.

The transport coefficient corresponding to $A$, which is denoted here as
$\eta_A$, comes from the Kubo formula \cite{LeBellac:1996}:
\begin{equation}
  \eta_A = \lim\limits_{k_0\to 0} \frac{C_A(k)}{k_0}.
\end{equation}
It reads with the current representation
\begin{eqnarray}
  \eta_A =\frac\beta Z \sum_{{\cal Q},{\cal P}}&&\pint4q
  \rh_{\cal Q}(q)\rh_{\cal P}(q)\,e^{-\beta q_0} 
  \left|\exv{q,{\cal Q}\left| A\right|q,{\cal 
        P}}\right|^2\nn&&
\end{eqnarray}
where we have omitted the zero argument of $A$.

Now we go to low temperatures, $\beta\to\infty$. The $\exp(-\beta q_0)$ factor
forces to keep only those quantum channels, which contain the lowest energy
levels while the matrix element is not zero there. For simplicity we assume,
that this is true for a single quantum channel, but this is not crucial for
the later discussion. Therefore in the followings we will suppress the notion
of the quantum channel, and write
\begin{equation}
  \eta_A =\frac\beta Z \pint4q \rh^2(q)\,e^{-\beta q_0} \left|\exv{q\left|
        A\right|q}\right|^2.
\end{equation}
Analogous result could be found for the viscosity in strong coupling
expansion at small temperatures \cite{Jakovac:2008ft}.

We should work out the low temperature expression for the entropy, too. The
free energy is the logarithm of $Z$, which can be written, by explicitely
separating the vacuum contribution:
\begin{equation}
  Z-1= \!\!\sum_{\ket n \neq\ket0}\!\!\exv{n\left|e^{-\beta H}\right|n} 
  = \!\!\sum_{{\cal Q}\neq\ket0}\!\!\pint4q\rh_{\cal Q}(q)\,e^{-\beta q_0}.
\end{equation}
At low temperatures the correction to 1 is small, and so we can approximate
$\ln (1+\delta Z) = \delta Z$ to leading order. We again assume that the most
important contributions come from a single quantum channel, and suppress the
index ${\cal Q}$ in the followings. It could also happen, that the relevant
quantum channel for entropy is different from the one for $\eta_A$, but we
will consider only those transport coefficients, where this is not the
case. Transport coefficients coming from the commutator of the energy-momentum
tensor (like the shear viscosity), are in this class, since all excitations
necessarily contribute to the energy-momentum tensor. Then at low temperatures
the free energy density and the entropy density read:
\begin{equation}
  f = -T \pint4q\,\rh(q)\,e^{-\beta q_0},\qquad
  s= -\frac{\d f}{\d T}
\end{equation}
At low temperatures, therefore, the ratio of $\eta_A$ and the entropy density
reads
\begin{equation}
  \frac{\eta_A}s = \frac{\displaystyle \beta \pint4q \rh^2(q) e^{-\beta q_0}
    \left|\exv{q\left|A\right|q}\right|^2}{\displaystyle \frac\d{\d T}
    \,T\pint4q\,\rh(q)\,e^{-\beta q_0}},
\end{equation}
where we have taken into account that at low temperatures
$Z\approx1$. 

In the following we make kinematical considerations to estimate the
temperature dependence of the $\eta/s$ ratio. There we should assume something
about the matrix element $\left|\exv{q\left|A\right|q}\right|^2$. Since we
calculate transport coefficients, we may assume that $A$ is a current, and
therefore its matrix element is zero, when there is no gradient of the
corresponding density. Therefore we assume that
\begin{equation}
  \left|\exv{q\left| A\right|q}\right|^2 = {\bf q}^2 {\cal A}.
\end{equation}
The coefficient ${\cal A}$ can be temperature dependent, but at small
temperatures we may approximate it with its zero temperature limit. One should
note that for the shear viscosity the current is a tensor, because we describe
the current of the energy-momentum four vector. But it does not alter the
statement that the matrix element remains to be proportional to $\q^2$ where
$\q$ describes the change of the momentum in the perpendicular direction.

So we can write:
\begin{equation}
\label{etaoverS_gen}
  \frac{\eta_A}s =\frac{ {\cal A} \beta \displaystyle \pint4q \rh^2(q) {\bf
      q}^2 e^{-\beta q_0}} {\displaystyle\frac\d{\d T} \,T \pint4q
    \,\rh(q)\,e^{-\beta q_0}}. 
\end{equation}
We will consider two different types of systems: one with a quasiparticle
behavior at the lowest energies, the other with a threshold behavior, and
examine the $\eta_A/s$ ratio.

\section{Quasiparticle case}

Here we assume a Breit-Wigner-type distribution of the lowest energy
eigenvalues:
\begin{equation}
  \rh(q) = \frac{2\Gamma}{(q_0-\ep_q)^2 + \Gamma^2},
\end{equation}
where $\ep_q$ is the dispersion relation of the quasiparticle, $\Gamma$ is the
quasiparticle width. Here we will assume a relativistic dispersion relation
$\ep_q^2 =q^2+m^2$. If $\Gamma\to0$ this form approaches
$2\pi\delta(q_0-\ep_q)$. We assume that we are in the ``small width'' regime,
which means that always, where we can, we should send $\Gamma\to0$.

This policy suggests that we should treat the product of two distribution
function in the sense of the Fermi's Golden Rule, ie. we treat one of them as
if it was a Dirac-delta:
\begin{equation}
\label{Fermirule}
  \rh^2(q) \approx \rh(q_0=\ep_q) \rh(q) = \frac2\Gamma\, \rh(q).
\end{equation}
Thereafter the integrals both in the numerator and in the denumerator contain
one spectral function, which can now be approximated by the delta
function. 

The free energy density reads
\begin{eqnarray}
\label{qpf}
  f&&= -T \pint4q \rh(q)\,e^{-\beta q_0} = -\frac{T}{2\pi^2}
  \int\limits_0^\infty \!dq\,q^2 e^{-\beta\ep_q} =\nn&&= -\frac{Tm^3}{2\pi^2}
  \int\limits_1^\infty\!dz\,z\sqrt{z^2-1} e^{-\beta m z} =
  -\frac{T^4}{2\pi^2} (\beta m)^2 K_2(\beta m),
\end{eqnarray}
where $K_2$ is the modified Bessel function of the second kind. At zero
mass (conformal case) the relevant limit is $\beta m\to 0$, then we find
\begin{equation}
  \label{qpconf}
  f|_{\mathrm{conf}} = -\frac{T^4}{\pi^2},\qquad s|_{\mathrm{conf}}
  = \frac{4T^3}{\pi^2}.
\end{equation}
In the massive case at temperatures $T\ll m$ we find in the leading order:
\begin{equation}
  f|_{\mathrm{m}} = -T^4 \left(\frac{\beta m}{2\pi}\right)^{3/2}\!\!
  e^{-\beta m},\quad s|_{\mathrm{m}} = \frac{m^{5/2}
    T^{1/2}}{(2\pi)^{3/2}} e^{-\beta m}.
\end{equation}

For the transport coefficient we find with help of \eqref{Fermirule}
\begin{eqnarray}
\label{qpeta}
  \eta_A &&= {\cal A}\beta \pint4q \rh^2(q)\q^2 e^{-\beta q_0}
  \approx \frac{2{\cal A}}{T\Gamma} \pint4q \rh(q) \q^2 e^{-\beta
    q_0} =\nn&&= \frac{{\cal A}m^5}{T\Gamma \pi^2} \int\limits_1^\infty \!dz\,
  z (z^2-1)^{3/2} e^{-\beta m z} = \frac{3{\cal A}m^3T}{\Gamma \pi^2}
  K_3(\beta m),
\end{eqnarray}
where $K_3$ is the modified Bessel function of the third kind. In the
conformal case:
\begin{equation}
  \eta_A|_{\mathrm{conf}}= \frac{24{\cal A}T^4}{\Gamma \pi^2}.
\end{equation}
In the massive case, at temperatures $T\ll m$ we find in the leading order:
\begin{equation}
  \eta_A|_{\mathrm{m}} =\frac{6{\cal A} }{\Gamma T}
  \frac{(mT)^{5/2}}{(2\pi)^{3/2}} e^{-\beta m}.
\end{equation}
Therefore the $\eta_A/s$ ratio reads in both cases
\begin{equation}
  \frac{\eta_A}s = 6{\cal A} \frac{T}{\Gamma}.
\end{equation}

As we can see, the desired ratio is proportional to $T/\Gamma$. This is the
manifestation of the qualitative argumentation of Ref. \cite{Kovtun:2004de}:
since $T$ is the kinetic energy and $1/\Gamma$ is the lifetime, therefore
$\eta_A/s \sim E_{\mathrm{kin}} \tau$. Therefore this ratio is constrained by
the Heisenberg relation, and we expect to have a lower bound. In fact, in the
conformal case $\Gamma$ must be proportional to the temperature (there is no
other scale), and the ratio is constant. In the massive case, on the other
hand, the width is exponentially small. This is because to form a width, the
particle must scatter on thermal states, but their abundance is $\sim
e^{-M/T}$, where $M$ is the energy of the lowest lying scattering
state. Therefore
\begin{equation}
  \frac{\eta_A}s\biggr|_{\mathrm{conformal}} \sim \mathrm{const.},\qquad
  \frac{\eta_A}s\biggr|_{\mathrm{massive}} \sim T e^{M/T}\;
  \stackrel{T\to0}{\longrightarrow}\; \infty.
\end{equation}

\section{Non-quasiparticle case}

The other big class of theories is when the low energy spectrum cannot be
described by a narrow quasiparticle peak, but instead there is a continuous
density of state. In this case it is worth to stay in the four-dimensional
formalism, and write the integral in 4D polar-coordinates. We should take into
account that, since the energy spectrum has a lower bound (for stability)
which can be chosen to be zero, the $q_0$ integral is restricted to positive
values. Moreover the integrand is spatially symmetric. Then we can use:
\begin{equation}
  \pint4q \Theta(q_0) \to  \frac1{4\pi^3}  \int\limits_0^\infty dq q^3
  \int\limits_0^\infty d\eta \sinh^2\eta.
\end{equation}
The expression for the free energy can be rewritten as
\begin{equation}
  f = -\frac{T}{4\pi^3} \int\limits_0^\infty\!dq q^3 \rh(q)
  \int\limits_0^\infty\! d\eta \sinh^2\eta e^{-\beta q \cosh\eta} =
  -\frac{T^2}{4\pi^3} \int\limits_0^\infty\! dq q^2 \rh(q) K_1(\beta q).
\end{equation}
For the transport coefficient we obtain
\begin{equation}
  \eta_A = \frac{{\cal A}\beta}{4\pi^3} \int\limits_0^\infty\!dq q^5 \rh^2(q)
  \int\limits_0^\infty\! d\eta \sinh^4\eta e^{-\beta q \cosh\eta} =
  -\frac{3{\cal A}}{4\pi^3} \int\limits_0^\infty\! dq q^5 \rh^2(q) K_2(\beta
  q).
\end{equation}

At $T\to0$ limit we will use the zero temperature limit of the spectral
density which is Lorentz-invariant. Since the modified Bessel functions all
decrease as $K(x)\to (\pi/2x)^{1/2} e^{-x}$, the $\beta q$ argument in the
above expressions enhances the lowest momentum part of the spectral function,
ie. the threshold region. Sufficiently close to the threshold the spectral
density can be approximated by a power law: so we will assume that the
spectral density has the form in the relevant regime as
\begin{equation}
  \rh(q) = C\Theta(q-M) (q-M)^w,
\end{equation}
where $M$ is the threshold value and $C$ is some constant.

If there is no mass gap, then the expressions for $f$ and $\eta_A$ can be
exactly performed, yielding for the entropy density
\begin{equation}
  s = \frac{C(5+w)}{4\pi^3}\,T^{4+w}\, 2^{2+w}\, \Gamma\left(
    \frac{5+w}2 \right) \Gamma\left(\frac{3+w}2 \right),
\end{equation}
and
\begin{equation}
  \eta_A = \frac{3 {\cal A}C^2} {4\pi^3}\,T^{5+2w}\, 4^{1+w}
  \Gamma(1+w) \Gamma(3+w).
\end{equation}

The transport coefficient to entropy ratio is therefore
\begin{equation}
  \frac{\eta_A}s = \alpha_w {\cal A}C\,T^{1+w}
\end{equation}
where $ \alpha_w = 3\;2^w\,\Gamma(1+w)\Gamma(3+w)/[ (5+w) \Gamma((5+w)/2)
\Gamma((3+w)/2)]$.

If $M\neq 0$ then at $T\ll M$ we can use the asymptotic form for the modified
Bessel functions, and obtain
\begin{equation}
  s = \sqrt{\frac\pi2}\,\frac{C\Gamma(w+1)} {4\pi^3}\, T^{w+3/2}M^{5/2}
  e^{-\beta M},
\end{equation}
and
\begin{equation}
  \eta_A  = \sqrt{\frac\pi2}\,
  \frac{3{\cal A}C^2\Gamma(2w+1)} {4\pi^3}  T^{5/2+2w}M^{5/2} e^{-\beta M}
\end{equation}
The transport coefficient over entropy ratio is therefore
\begin{equation}
  \frac{\eta_A}s = \bar\alpha_w {\cal A}C\,T^{1+w},
\end{equation}
where $\bar\alpha_w =3\Gamma(1+2w)/\Gamma(1+w)$.

So finally from both the gapped and gapless case we obtained the same result:
\begin{equation}
  \frac{\eta_A}s \sim T^{1+w} \stackrel{T\to0}{\longrightarrow}\; 0
\end{equation}
for any realistic threshold behavior ($w>-1$).

We remark here that the result could be guessed by dimensional argumentation.
The coefficient $C$ must be dimensionfull: $\rh$ is of dimension $1/E$, $q,M
\sim E$ therfore $C\sim E^{-1-w}$. Since $\eta_A$ contains $C^2$, the entropy
density $C$, their ratio is $\sim C$. But the ratio is dimensionless, so a
quantity with dimension $E^{1+w}$ must appear. If $M=0$ the only candidate is
$T$, so we must have $T^{1+w}$. If $M\neq0$, in principle there could be also
some $M$ factor, but it turns out from the concrete calculation, that the
$M$ factors drop out.

\section{Conclusion}

For conclusion we recall that we studied the low temperature behavior of the
ratio of some transport coefficient $\eta_A$ (where the notation is motivated
by the shear viscosity) and the entropy density. The computation is based on
exact formulae which are used near zero temperature. We studied two classes of
models, both describing relativistic superfluids.

The first is the class of the Bose-liquids, where the lowest energy excitation
is a stable particle at zero temperature which becomes a narrow width
quasiparticle at small temperatures. To this class belong a lot of particle
physics models, for example the $\Phi^4$ model, or the sigma model. There we
have found that the $\eta_A/s$ ratio is proportional to $T/\Gamma$, in
agreement with the qualitative expectations \cite{Kovtun:2004de}. In the
conformal limit this ratio is constant, in the massive case the ratio is
diverging as $T\to0$. In this class therefore there is a lower bound for the
$\eta_A/s$: in the case of viscosity it is probably $1/4\pi$
\cite{Policastro:2001yc}.

The second class of the studied models is the non-Bose-liquids: here the
lowest lying excitations are form a continuous spectrum. The pure Yang-Mills
theories may belong to this class, according to the strong coupling expansion
\cite{Jakovac:2008ft}. The result in these models is that $\eta_A/s \sim
T^{1+w}$ where $w$ is the power of the power law appearing in the density of
states when expanded around the threshold. In generic case this ratio goes to
zero at zero temperature. Therefore in these class of models there is no lower
bound for the shear viscosity entropy density ratio.

\section*{Acknowledgments}

The author acknowledges discussions with D. Nogradi. He also
acknowledges support from the Humboldt Foundation and the Hungarian
Science Fund (OTKA) under grant K68108.

\end{document}